\documentclass[final,5p,times,twocolumn, number, sort&compress]{elsarticle}

\usepackage{amssymb}
\usepackage{amsmath}
\usepackage{graphicx}
\usepackage{subcaption}
\usepackage{float}
\usepackage[colorlinks=true]{hyperref}
\usepackage{appendix}
\usepackage{amsbsy}
\usepackage{slashed}
\usepackage{tikz}
\usepackage{tikz-feynman}
\usepackage{kantlipsum}
\usepackage{cuted}
\DeclareMathAlphabet\mathbfcal{OMS}{cmsy}{b}{n}

\newcommand{\code}[1]{\textsc{#1}}

\journal{Physics Letters B}

\begin{document}
\begin{frontmatter}
\title{\boldmath Planar three-loop QCD helicity amplitudes for $V$+jet production at hadron colliders}


\author[first]{Thomas Gehrmann}            
\author[first]{Petr Jakub\v{c}\'{i}k}     
\affiliation[first]{organization={Physik-Institut, Universität  Zürich}, 
            addressline={Winterthurerstrasse 190,CH-8057 Zürich},
            country={Switzerland}} 

\author[second]{Cesare Carlo Mella}            
\author[second]{Nikolaos Syrrakos}
\author[second]{Lorenzo Tancredi}
\affiliation[second]{organization={Technical University of Munich, TUM School of Natural Sciences, Physics Department},
addressline={James-Franck-Straße 1, 85748 Garching},
country={Germany}}

\begin{abstract}
We compute the planar three-loop Quantum Chromodynamics (QCD) corrections to the helicity amplitudes involving a 
vector boson $V=Z,W^\pm,\gamma^*$, two quarks and a gluon. These amplitudes are 
relevant to vector-boson-plus-jet production at hadron 
colliders and other precision QCD observables. The 
planar corrections encompass the leading colour factors $N^3$, $N^2 N_f$, $N N_f^2$ and $N_f^3$.
 We provide the finite remainders of the independent helicity amplitudes in terms of multiple polylogarithms, continued to all kinematic regions and in a form which is compact and lends itself to efficient numerical evaluation. The presented amplitude respects the conjectured symbol-adjacency constraints for amplitudes with three massless and one massive leg.
\end{abstract}

\end{frontmatter}

\section{Introduction}\label{sec:intro}
Scattering amplitudes provide the connection between the Lagrangian formulation of the Standard Model of particle physics and observable quantities at particle collider experiments. 
Their perturbation theory expansion amounts to the computation of loop corrections, enabling increasingly accurate predictions.  
Multi-loop amplitudes in QCD provide moreover an 
important gateway for new mathematical ideas to enter particle physics. 

The helicity amplitudes for processes involving a vector boson and three partons have been of pivotal
importance to QCD precision physics. In their different 
kinematical crossings, these amplitudes describe 
three-jet production in $e^+e^-$ 
annihilation~\cite{Ellis:1976uc}, (2+1)-jet production in deep inelastic $ep$ scattering, and $V$+jet production at hadron colliders. Measurements of all 
three types of processes allowed establishing
QCD as the correct theory of the strong interactions. Precision data on these benchmark reactions continue to play a key role in enabling accurate determinations of the QCD coupling constant and of the partonic content of the proton. 

On the theoretical side, these processes are also 
a testing ground for the development of new 
calculational methods in perturbative QCD. $e^+e^- \to 3$~jets was the first jet production 
process to be computed to 
next-to-leading order (NLO,~\cite{Ellis:1980wv}) and next-to-next-to-leading order (NNLO,~\cite{Gehrmann-DeRidder:2007vsv,Weinzierl:2008iv}), 
thereby driving the development of 
 systematic formulations of both subtraction and phase-space slicing methods for handling infrared singularities and 
 their implementation in 
 flexible parton-level event generation programs~\cite{Kunszt:1989km,Giele:1991vf,Catani:1996jh,Gehrmann-DeRidder:2014hxk}.  
It is likely that $e^+e^- \to$ 3 jets will be among the first processes with jets to be tackled at
third order (N$^3$LO) in perturbative QCD.

At the LHC, leptonic decays of produced vector bosons leave clear signatures in the detectors, allowing observables such as the 
transverse momentum distribution of the lepton pair to be measured to a precision well below the percent level. 
To match this level of accuracy requires going beyond
the currently available NNLO predictions~\cite{Boughezal:2015dva,Gehrmann-DeRidder:2015wbt,Neumann:2022lft} for 
vector-boson-plus-jet production, which
relates directly 
to the vector boson transverse momentum spectrum. 

The Born-level helicity amplitudes for these processes 
couple the vector boson to a quark-antiquark pair and a gluon: $Vq\bar qg$. QCD corrections to these amplitudes were computed previously up to one loop~\cite{Giele:1991vf} and two loops~\cite{Garland:2002ak}, recently extended to
include singlet axial vector coupling 
terms~\cite{Gehrmann:2022vuk} and higher orders 
in the dimensional regulator~\cite{Gehrmann:2023zpz}. 

The extension of scattering amplitudes
to the next loop order is always associated with conceptual advances. More than twenty years ago, the calculation of the two-loop master integrals for four-point functions with one off-shell leg~\cite{Gehrmann:2000zt, Gehrmann:2001ck} served as the first advanced example of the use of the method of differential equations~\cite{Kotikov:1990kg, Gehrmann:1999as}, now ubiquitous in multi-loop calculations. It also necessitated the study of a new class of functions, generalized harmonic polylogarithms (multiple polylogarithms, MPLs)~\cite{Gehrmann:2000zt,Goncharov:1998kja}. 

The concept of a canonical basis of master integrals~\cite{ArkaniHamed:2010gh,Henn:2013nsa} has facilitated the extension of the two-loop $Vq\bar qg$
master integrals to transcendental weight six~\cite{Gehrmann:2023etk}. At three loops, the ladder-box topology was considered some time ago~\cite{DiVita:2014pza}, while the more challenging tennis court topologies have only been addressed more recently~\cite{Canko:2021xmn, Canko:2023yoe}. Interestingly, all planar integrals can be expressed in terms of MPLs with the same alphabet as at two loops. On the other hand, the first results for non-planar topologies~\cite{Henn:2023vbd, Canko:2023yoe} indicate an extended alphabet 
with the appearance of square roots. 
The increasing complexity at intermediate stages of amplitude calculations has also led to the adoption of methods based on finite-field arithmetic in particle physics~\cite{vonManteuffel:2014ixa,Peraro:2019svx}, allowing the reconstruction of analytic results for the amplitudes from 
multiple numerical evaluations. 

This combination of conceptual and technical developments enables us to 
derive the planar three-loop QCD $Vq\bar qg$ helicity amplitudes,  which we present in this letter. These amplitudes  constitute the  virtual N$^3$LO QCD corrections to $e^+e^- \to$ 3 jets and processes related to it by kinematical crossing, and are the stepping stone to reach a new level of precision in QCD studies at colliders.

\section{Setup and Calculation}\label{sec:setup}
We consider the production of a vector boson $V$ through lepton-antilepton annihilation and its subsequent decay into a quark, anti-quark and a gluon
\begin{equation}
l^{+}(p_5)+l^{-}(p_6) \to V(p_4) \rightarrow q(p_1) + \bar{q}(p_2) + g(p_3)\, .
\end{equation}
The calculation is performed assuming a purely vector coupling for $V$ and the effect of non-singlet axial couplings is reconstructed later by reweighting the
helicity amplitudes by the appropriate couplings, see eq.~\eqref{eq:HALpL}.
We introduce the usual Mandelstam invariants
\begin{align}\nonumber
s_{12} = (p_1+p_2)^2\,, \quad   s_{13} = (p_1+p_3)^2\,, \quad  s_{23} = (p_2+p_3)^2
\end{align}
which satisfy the conservation equation
\begin{equation}
s_{12} + s_{13} + s_{23} = q^2\,,
\label{eqn:mandelstamsum}
\end{equation}
where $q^2=p_4^2$ is the momentum-squared of the vector boson. It is convenient to work with the following dimensionless ratios
\begin{align}
    x = {s_{12}}/{q^2}\,, \quad \quad  y = {s_{13}}/{q^2}\,, \quad \quad z = {s_{23}}/{q^2}\,, 
    \label{eqn:xyz}
\end{align}
such that \eqref{eqn:mandelstamsum} implies $x+y+z=1$. It is easy to see that in the decay kinematic region all these invariants are non-negative and
\begin{equation}
 z \geq 0\,, \quad \quad  0 \leq  y  \leq 1 - z\,, \quad  \quad x = 1 - y -z\,. 
\end{equation}

We work in dimensional regularization, taking ${d = 4 -2 \epsilon}$. In particular, we adopt the 't Hooft-Veltman dimensional regularization scheme~\cite{tHooft:1972tcz}, in which loop momenta are $d-$dimensional but external states are kept in four dimensions. The amplitude for the production of a vector boson can then be written as 
\begin{align}
\mathcal{M}_{} &= -i\sqrt{4\pi\bar{\alpha}_s}\, {\mathbb T_{ij}^{a}} A_{ \mu \nu} \epsilon_3^{\mu} \epsilon_4^{\nu}\,,
\label{eq:Ampl}
\end{align}
where $\bar{\alpha}_s$ is the bare strong coupling and ${\mathbb T_{ij}^a}$ is the generator of $SU(N)$ in the fundamental representation.
We can then proceed by decomposing the amplitude $A^{\mu\nu}$ in eq.~\eqref{eq:Ampl} into a basis of tensor structures in $d=4$ spacetime dimensions following a method introduced in~\cite{Peraro:2019cjj,Peraro:2020sfm}, see ref.~\cite{Gehrmann:2023zpz} for details.

Instead of calculating the corresponding form factors, we prefer to start from this decomposition 
and introduce projector operators that directly extract the amplitude for fixed helicities of the external states and contracted with a leptonic current.
To this end, we define
\begin{equation}
    {\rm M}_{\lambda_{q_2}\lambda_3 \lambda_{l_5}} = \epsilon_{3,\mu}^{\lambda_3}   \, 
	A^{\mu \nu}_{\lambda_{\bar q_1}\lambda_{q_2} } C_{\lambda_{l_5}}^\nu(p_5,p_6)\,.
	 \label{eq:HAnocouplings}
\end{equation}
Working in spinor-helicity formalism, we write for the left- and right-handed fermionic currents 
\begin{align}\label{eq:leptonspinhel1}
    C_L^\mu(p,q) = [q \gamma^\mu p\rangle\,, \qquad C_R^\mu(p,q) = [p \gamma^\mu q\rangle\,, 
\end{align}
and for the polarization vector of the outgoing external gluon 
\begin{equation}\label{eq:gluonpol}
\epsilon_{3,-}^\mu = \frac{\langle 3 \gamma^\mu 2 ]}{\sqrt{2} [3 2]}\,, \quad 
\epsilon_{3,+}^\mu = \frac{\langle 2 \gamma^\mu 3 ]}{\sqrt{2} \langle 2 3 \rangle}\,,
\end{equation}
where we adopted an axial gauge with $p_2$ as the reference vector.
In this setup, the two independent helicity amplitudes read
\begin{alignat}{3}
    {\rm M}_{L+L} &= \frac{1}{\sqrt{2}}  
	 \Big[ \langle 1 2 \rangle [1 3]^2 \Big( \alpha_1 \langle 536] 
	&&+ \alpha_2 \langle 526] \Big)\nonumber\\
        &{}&&+ \alpha_3 \langle 25 \rangle [13] [36] \Big]\,,
\label{eq:mLLv}\\
	 {\rm M}_{L-L} &=  \frac{1}{\sqrt{2}}  
	 \Big[ \langle 23 \rangle^2 [1 2] \Big( \gamma_1 \langle 536] 
	&&+ \gamma_2 \langle 516] \Big)\nonumber\\        
        &{}&&+ \gamma_3 \langle 23 \rangle \langle 35 \rangle [16] \Big], \label{eq:LpLv}
\end{alignat}
and the remaining two helicity amplitudes, ${\rm M}_{R+L}$ and ${\rm M}_{R-L}$, can be obtained by means of parity and charge conjugation~\cite{Gehrmann:2022vuk,Gehrmann:2023zpz}.
Additionally, for each amplitude, the lepton current handedness can be reversed by a simple swap $p_5 \leftrightarrow p_6$.
The contribution of a Feynman diagram to the independent helicity coefficients $\alpha_i$ and $\gamma_i$ with $i=1,...,3$ can then be obtained by applying suitable projector operators. Their explicit form (immaterial to the present discussion) is given in the supplemental material.

\begin{figure}[tp]%
      \begin{minipage}{.42\columnwidth}
      \raggedright
          \includegraphics[width=0.95\linewidth]{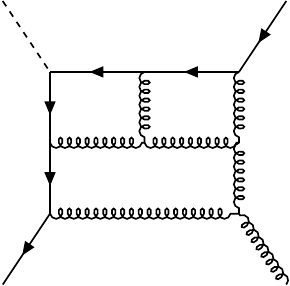}
          \subcaption{$N^3$}
      \end{minipage}%
      \begin{minipage}{.42\columnwidth}
      \raggedleft
          \includegraphics[width=0.95\linewidth]{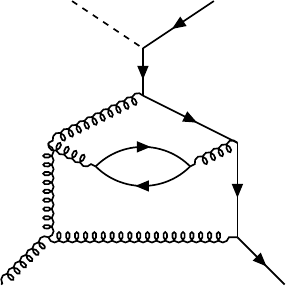}
          \subcaption{$N^2 N_f$}
      \end{minipage}%

\bigskip
      \begin{minipage}{.98\columnwidth}
          \centering
          \includegraphics[width=0.54\linewidth]{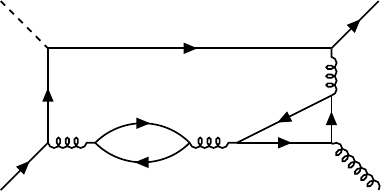}
          \subcaption{$N N_f^2$}
      \end{minipage}
\caption{ Representative three-loop planar diagrams which contribute to the three leading color layers.}\label{fig:representative_diags}
\end{figure}

The helicity amplitude coefficients 
$\Omega = (\alpha_i,\gamma_i)$ can be expanded in powers of the  strong coupling constant
\begin{align}
\Omega &= \Omega^{(0)} + \left( \frac{\alpha_s}{2 \pi}\right) \Omega^{(1)}
     + \left( \frac{{\alpha}_s}{2 \pi}\right)^2 \Omega^{(2)} \nonumber \\
    &+ \left( \frac{{\alpha}_s}{2 \pi}\right)^3 \Omega^{(3)} +\mathcal{O}({\alpha}_s^4) \,.   \label{eq:omegaexp}
\end{align}
This expansion holds for the bare coefficients (with $\alpha_s$ replaced by $\bar{\alpha}_s$), the renormalized coefficients, as well as for the IR-subtracted coefficients defined below. 
The three-loop coefficients 
are in turn decomposed into color factors as
\begin{align}\label{eq:colordecomp}
\Omega^{(3)}&=   
N^3\:\Omega^{(3)}_1
+ N \Omega^{(3)}_2
+ \frac{1}{N} \Omega^{(3)}_3
+ \frac{1}{N^{3}} \Omega^{(3)}_4\nonumber\\
&+ N_{f}  \:N^2 \Omega^{(3)}_5
+ N_{f} \Omega^{(3)}_6
+ \frac{N_{f}}{N^{2}} \Omega^{(3)}_7\nonumber\\
&+ N_{f}^2  N \:\Omega^{(3)}_8
+ \frac{N_{f}^2}{N} \Omega^{(3)}_9 
+ N_{f}^3 \Omega_{10}^{(3)}
\nonumber\\
&+ N_{f,V}  N^2 \Omega^{(3)}_{11}
+ N_{f,V} \Omega^{(3)}_{12}\nonumber\\
&+ \frac{N_{f,V}}{ N^{2}} \Omega^{(3)}_{13}
+ N_{f} N_{f,V}  N \Omega^{(3)}_{14}
+ \frac{N_{f} N_{f,V}}{ N} \Omega^{(3)}_{15}, 
\end{align}
where $N_f$ is the number of active quark flavors in the loops, and $N_{f, V}$ denotes contributions where the vector boson couples to an internal fermion loop. In this letter we focus on 
the dominant terms in this expansion. 
Using $N_f \sim N$, these are all the terms with total power of three in $N$ and $N_f$
and they correspond to $\Omega_j$ with $j=1,5,8,10$.
Importantly, the leading color factors only receive contribution from planar diagrams, see Figure~\ref{fig:representative_diags} for example diagrams. 
The contributions proportional to $N_f^2 N_{f,V}$ vanish for a vector-like coupling by Furry's theorem. 
Eq.~\eqref{eq:colordecomp} is also valid for any helicity amplitude coefficient -- bare, renormalized or IR subtracted. 

In the following, we fix the renormalization scale in 
$\Omega$ as $\mu^2=q^2$.
The full scale dependence can then be recovered through
\begin{align}
     \Omega^{(3)}(\mu) & = 
     \left(\frac{5}{16} \beta _0^3 L(\mu)^3+\beta _0 \beta _1 L(\mu)^2+\frac{1}{2} \beta _2 L(\mu)\right)\Omega^{(0)}\nonumber\\
    &+  \left(\frac{15}{8} \beta _0^2 L(\mu)^2+\frac{3}{2} \beta _1 L(\mu)\right) \Omega^{(1)}\nonumber\\
   &     +\frac{5}{2} \beta _0 L(\mu) \Omega ^{(2)} + \Omega ^{(3)} 
\end{align}
with $L(\mu) = \log \left(\mu ^2/q^2\right)$.

The helicity amplitudes for the decay of a Standard Model vector boson $V$ can finally be related to the helicity amplitudes obtained above by dressing with the appropriate electroweak couplings
\begin{align}
\mathcal{M}_{\lambda_{q_2}\lambda_3 \lambda_{l_5}}^V =& - 
 \frac{i \sqrt{4 \pi \alpha_s} (4 \pi \alpha)\, L_{l_5 l_6}^{V} L_{q_1 q_2}^V }{D(p_{56}^2,m_V^2)} \nonumber \\ 
 &\times
 {\mathbb T^{a}_{ij}} \,{\rm M}_{\lambda_{q_2}\lambda_3 \lambda_{l_5}} \,, \label{eq:HALpL}
 \end{align}
where $p_{56} = p_5 + p_6$, the vector boson propagator reads
\begin{align}
D\left(q^2,m_V^2\right) &= q^2 - m_V^2 + i \Gamma_V m_V
\end{align}
and the couplings for the bosons ${V=Z,W^\pm,\gamma^*}$ are
\begin{align}
    R_{f_1 f_2}^\gamma = L_{f_1 f_2}^\gamma &= - e_{f_1} \delta_{f_1 f_2}\,,\\
    L_{f_1 f_2}^Z &= \frac{I_3^{f_1} - \sin^2{\theta_w} e_{f_1}}{\sin{\theta_w} \cos{\theta_w}} \delta_{f_1 f_2} \,, \\
    L_{f_1 f_2}^W &= \frac{\epsilon_{f_1,f_2}}{\sqrt{2}\sin{\theta_w}}  \,,\\
    R_{f_1 f_2}^Z &=-\frac{\sin{\theta_w} e_{f_1}}{\cos{\theta_w}} \delta_{f_1 f_2}\,,\\
    R_{f_1 f_2}^W &=0  \,.
\end{align}  
In the formulas above, $\alpha$ is the electroweak coupling constant,
$\theta_w$ is the Weinberg angle, $I_3 = \pm 1/2$ is the third component
of the weak isospin and the charges $e_i$
are measured in terms of the fundamental electric charge $e>0$.
Moreover, $\epsilon_{f_1,f_2}=1$ if $f_1 \neq f_2$ but belonging to the same
isospin doublet, and zero otherwise. 

In order to compute the (unrenormalized) corrections to the helicity amplitude coefficient, we use the same unified workflow as for the tree-level, one- and two-loop amplitudes for $Vq\bar{q}g$~\cite{Gehrmann:2023zpz}, whose agreement with older results in the literature up to the finite part in $\epsilon$ provides an additional check on our method. In summary, the relevant three-loop diagrams are  generated using \code{QGRAF}~\cite{Nogueira:1991ex} and every manipulation including insertion of Feynman rules, evaluation of  Dirac and Lorentz algebra and application of the projectors are performed in \code{FORM}~\cite{Vermaseren:2000nd}. 
Once the helicity projectors have been applied, all Feynman diagrams are expressed in terms of scalar integrals, which can be written in terms of a single planar auxiliary topology of the form
\begin{equation}
\mathcal{I}_{n_1,...,n_{15}} = e^{3 \gamma_E \epsilon} \int \prod_{i=1}^3 \frac{d^d k_i}{i \pi^{d/2}} \frac{1}{D_1^{n_1}... D_{15}^{n_{15}}}
\end{equation}
with $\gamma_E=0.5772\ldots$ the Euler constant and propagators
\begin{align*}
\begin{tabular}{l@{\hskip 0.2in}l@{\hskip 0.2in}l}
$D_1 = k_1 $    & $D_6 = k_3 - p_1$               & $D_{11} = k_2 - p_{123}$ \\
$D_2 =  k_2$    & $D_7 = k_1 - p_{12}$          & $D_{12 }= k_3 - p_{123}$  \\
$D_3 = k_3 $    & $D_8 = k_2 - p_{12} $         & $D_{13} = k_1 - k_2 $           \\
$D_4 = k_1-p_1$ & $D_9 = k_3 - p_{12} $         & $D_{14} = k_1 - k_3$            \\
$D_5 = k_2-p_1$ & $D_{10} = k_1 - p_{123}$ & $D_{15} = k_2 - k_3$          
\end{tabular}
\end{align*}
with $p_{ij(k)} = p_i + p_j (+p_k)$.
The integrals can be reduced to a set of master integrals using integration-by-parts (IBP)
identities~\cite{Tkachov:1981wb,Chetyrkin:1981qh}. For the actual reduction, 
we use the implementation of the Laporta
algorithm~\cite{Laporta:2001dd} in the automated code \code{Kira2}~\cite{Maierhoefer:2017hyi,Klappert:2020nbg} and express all integrals
directly in terms of the canonical basis for the three-loop planar family defined in~\cite{Canko:2023yoe}.
Here it was shown that, in line with the one- and two-loop results, the three-loop planar integrals can be evaluated to arbitrary orders in the dimensional regularization
parameter $\epsilon$ in terms of multiple polylogarithms (MPLs)~\cite{Goncharov:1998kja,Remiddi:1999ew,Gehrmann:2000zt,Vollinga:2004sn}
with alphabet
$\{y,z,1-y,1-z,y+z,1-y-z\}$.

The amplitude before reduction can be expressed in terms of 95625 scalar integrals, which in turn are reduced to 291 canonical basis elements and their crossings. 
The size and complexity of intermediate expressions makes the use of traditional methods for symbolic insertion of the IBP reduction into the unreduced amplitude highly non-trivial. 
Therefore, 
 in view of the expected increase in complexity of the subleading layers in the color expansion~\eqref{eq:colordecomp}, 
 we also devised a hybrid method involving finite field reconstruction, in parallel to a standard fully analytic approach. 
 
 In particular, in the standard approach, we produced the IBP identities with~\code{Kira2} and used~\code{Mathematica} 
 and~\code{Fermat} to include them into the unreduced amplitude and simplify the resulting coefficients. We then expanded the master integrals in $\epsilon$ and obtained the final expression for the amplitude in terms of MPLs.
 In the hybrid approach, we used the same IBP reduction obtained with \code{Kira2}. However, we then performed an insertion of the reduction and MI solutions in the amplitude for a handful of rational numerical values of the variables $y$ and $z$ in the coefficients of the MPLs. This step allowed us to determine which polylogarithms appear in the final expression for a given helicity coefficient. With this information, we proceeded to reconstruct the rational prefactors in $y$ and $z$ only for the relevant polylogarithms, leveraging the implementation provided by \code{FiniteFlow}~\cite{Peraro:2019svx}. This approach lends itself to parallelization and gives us the possibility to recycle the symbolic reductions in future calculations. Compared to the traditional approach of reconstructing only the finite part, it allows us to present also the renormalized amplitudes with explicit IR poles for greater versatility in phenomenological applications with different pole schemes. We verified that the reconstructed results exactly matched those obtained through the fully analytic approach.
\section{UV renormalization and IR subtraction}\label{sec:UVandIR}
All UV divergences are removed by expressing the helicity coefficients in terms of the
$\overline{\rm MS}$-renormalized strong coupling $\alpha_s=\alpha_s(\mu)$.
Since this is a standard procedure, we only summarize the necessary $\beta$-function coefficients in the supplemental material.
The remaining IR divergences are universal and can be extracted multiplicatively 
in the formalism of Soft-Collinear Effective Theory (SCET)~\cite{Becher:2009qa, Catani:1998bh, Dixon:2009ur,Gardi:2009qi} as
\begin{equation}
    \mathbf{\mathbf{\Phi}}_{\text{finite}}(\{p\}) = \lim_{\epsilon \rightarrow 0} \mathbfcal{Z}^{-1}(\{p\}; \epsilon) \,\, \mathbf{\mathbf{\Phi}}({p}, \epsilon)\,,
    \label{eqn: IR-Renormalization}
\end{equation}
where we indicate with $\mathbf{\mathbf{\Phi}}_{\text{finite}}(\{p\})$ the finite remainder of the amplitude $\mathbf{\mathbf{\Phi}}({p}, \epsilon)$. As indicated by
the bold font, 
$\mathbfcal Z$ is generally an operator in color space, but it diagonalizes when acting on the $Vq\bar q g$ amplitude coefficients, which factorize onto 
a single color structure,
\begin{equation}
    \mathbf{\Phi}={\mathbb T^a_{ij}}\,\Omega\,.
    \label{eq:colvec}
\end{equation} 
$\mathbfcal Z$ can be formally expressed in terms of the anomalous dimension matrix $\mathbf\Gamma$ as
\begin{align}\nonumber
    \mathbfcal{Z}(\epsilon, \{p\}, \mu) &= \mathbb{P} \,  \text{exp} \Big[  \int_{\mu}^{\infty} \frac{d\mu'}{\mu'} \mathbf{\Gamma}(\{p\}, \mu') \Big] \\
    &= \sum_{l = 0}^{\infty} \Big( \frac{\alpha_s}{2 \pi} \Big)^l \mathbfcal{Z}^{(l)} \,.
    \label{eqn: Z_operator}
\end{align}

Importantly, for the first time at three loops, the anomalous dimension operator is not entirely captured by a dipole term but receives contributions also from quadrupole color correlation operators,
\begin{equation}\label{eq:dipole_+_quadrupole}
\mathbf{\Gamma}(\{p\}, \mu)=  \mathbf{\Gamma}_\text{dip}(\{p\}, \mu)  + \mathbf{\Delta}_3(\{p\}, \mu)  \; .
\end{equation}
The former is due to the pairwise exchange of color charge between 
external legs and reads
\begin{align}
 \mathbf{\Gamma}_{\text{dip}}(\{p\},\mu) &= \sum_{1 \leq i < j \leq n} \mathbf{T}_i^a\mathbf{T}_j^a \, \gamma^{\text{K}}(\alpha_s) \log\Big(-\frac{\mu^2}{s_{ij} + i \eta }\Big)\nonumber\\
 &\phantom{\quad}+  \, \sum_{i=1}^n \gamma^i(\alpha_s)\,,
 \label{DipoleExpression}
\end{align}
where the $\gamma^{\text{K}}$ is the cusp anomalous dimension and $\gamma^i$ is the  anomalous dimension of parton $i$. 

The quadrupole contribution $\mathbf{\Delta}_3$ in
eq.~\eqref{eq:dipole_+_quadrupole} accounts for the exchange
of color charge among three or four soft gluons emitted from (a subset of) the partonic legs. Unlike in three-loop four-parton scattering~\cite{Henn:2016jdu,Caola:2021rqz,Caola:2021izf,Caola:2022dfa} where this term was observed in N=4 SYM and QCD amplitudes for the first time, soft gluon emissions from four external partons are impossible here and thus the quadrupole term consists only of the entirely kinematics-independent term~\cite{Almelid:2015jia}
\begin{multline} \label{eq:quadrupole}
\mathbf{\Delta}^{(3), ijk}_3 = -f_{abe} f_{cde}\,
\bigg[
16 \,C   \, \sum_{i=1}^3 \sum_{\substack{1\leq j < k \leq3 \\ j,k\neq i}}  \left\{ \mathbf{T}^a_i,\mathbf{T}^d_i \right\} \mathbf{T}^b_j \mathbf{T}^c_k 
\bigg]\,,
\end{multline}
with  $C = \zeta_5 + 2 \zeta_2 \zeta_3$ a constant. However, this term only enters the color layers that are suppressed by at least $N^{-2}$ relative to the 
leading color contributions considered here.

\begin{table*}[t]
\begin{tabular}{|c||c|c|c|c|}
\hline
{} &  $V\to gq\bar{q}$ & $q\bar{q}\to Vg$ & $qg\to Vq$ & $\bar{q}g\to V\bar{q}$  \\ \hline
 {} & $y=z=1/3$ & \multicolumn{3}{c|}{$u=1/7, v=3/5$}  \\ \hline\hline
$\alpha_1$ & $0.59426734 + 0.03099861 i$ & $ -0.49538117 +    0.65353908 i$ & $ -0.22197101 - 0.03494388 i$ & $ -0.22197101 -    0.03494388 i$ \\ \hline
$\alpha_2$ & $ 1.7192645 - 8.7893957 i$ & $   2.0901940 + 8.1000376 i$ & $ -1.6429881 + 0.5830600 i$ & $ -1.6429881 +    0.5830600 i$ \\ \hline
$\alpha_3$ & $ 1.2128210 + 5.9479256 i$ & $ -6.0948209 - 9.3383011 i$ & $   0.94229209 - 0.78134227 i$ & $   0.94229209 - 0.78134227 i$ \\ \hline
$\gamma_1$ & $ -0.59426734 - 0.03099861 i$ & $   0.46544666 - 1.05443048 i$ & $ -0.05750758 +    0.15110700 i$ & $ -0.05750758 + 0.15110700 i$ \\ \hline
$\gamma_2$ & $ -1.7192645 +    8.7893957 i$ & $ -1.0680269 - 9.5317293 i$ & $ 0.24066096 - 0.32042746 i$ & $   0.24066096 - 0.32042746 i$ \\ \hline
$\gamma_3$ & $ 1.2128210 + 5.9479256 i$ & $ -5.2865869 -    7.5133951 i$ & $ 0.18231060 + 0.07136413 i$ & $ 0.18231060 + 0.07136413 i$ \\ \hline
\end{tabular}
\caption{Numerical values for the finite remainders of the helicity coefficients at a phase-space point in each channel to 8 significant figures, setting $q^2=1$, $N=3$ and $N_f=5$. All values are to be multiplied by $10^4$.}
\label{tab: numerics}
\end{table*}

In summary, given the $l$-loop renormalized helicity amplitudes $\mathbf{\Phi}^{(l)}$, one can remove the IR singularities with
\begin{align}
\mathbf{\Phi}^{(0)}_{\text{finite}} & = \mathbf{\Phi}^{(0)}, \nonumber \\
\mathbf{\Phi}^{(1)}_{\text{finite}} & = \mathbf{\Phi}^{(1)} - \mathbf{I}^{(1)} \, \mathbf{\Phi}^{(0)} ,
\nonumber \\
\mathbf{\Phi}^{(2)}_{\text{finite}} & = \mathbf{\Phi}^{(2)} - \mathbf{I}^{(2)} \, \mathbf{\Phi}^{(0)}  - \mathbf{I}^{(1)} \, \mathbf{\Phi}^{(1)}\,,
\nonumber \\
\mathbf{\Phi}^{(3)}_{\text{finite}} & = \mathbf{\Phi}^{(3)} - \mathbf{I}^{(3)} \, \mathbf{\Phi}^{(0)}  - \mathbf{I}^{(2)} \, \mathbf{\Phi}^{(1)} - \mathbf{I}^{(1)} \, \mathbf{\Phi}^{(2)}\,.
\label{eqn: perturbative IR subtraction three loops}
\end{align}
The subtraction operators in \eqref{eqn: perturbative IR subtraction three loops} can be expressed as
\begin{align}
 \mathbf{I}^{(1)} & = \mathbfcal{Z}^{(1)}\,,\nonumber \\
 \mathbf{I}^{(2)} & = \mathbfcal{Z}^{(2)} - \big(\mathbfcal{Z}^{(1)}\big)^2\,,\nonumber \\
 \mathbf{I}^{(3)} & = \mathbfcal{Z}^{(3)} - 2\mathbfcal{Z}^{(2)}\mathbfcal{Z}^{(1)} + \big(\mathbfcal{Z}^{(1)}\big)^3\,.
 \label{eqn: IR scet}
\end{align}
After performing the color algebra in (\ref{eqn: perturbative IR subtraction three loops}), we recover the simple factorizing structure (\ref{eq:colvec}) for 
the finite parts of the helicity amplitude. Consequently, 
(\ref{eqn: perturbative IR subtraction three loops})
also holds for the amplitude coefficients $\Omega$. 

As explained in detail in~\cite{Gehrmann:2023etk}, the explicit form of the operators $\mathbfcal{Z}_l$ in terms of the universal constants in the perturbative expansion for $\gamma^K$, $\gamma^q$ and $\gamma^g$ is implied by the dipole formula \eqref{DipoleExpression} and the new term $\mathbf{\Delta}^{(3)}_3$. A summary of the relevant expressions for UV renormalization and IR subtraction is given in the supplemental material.

\section{Results}\label{sec:helamps}
In view of future phenomenological applications 
of the three-loop $Vq\bar qg$ amplitudes to LHC physics, we perform the analytic continuation of the $V$ decay amplitudes to the kinematic regions where the electroweak boson is produced in addition to a jet from the scattering of two partons: $ q g \rightarrow V q$, $\bar{q} g \rightarrow V \bar{q}$ and $q \bar{q} \rightarrow V g $.

The general strategy for performing the analytic continuation of the MPLs appearing in $2\to 2$ scattering involving four-point functions with one off-shell leg and massless propagators was outlined in  detail in~\cite{Gehrmann:2002zr} and adapted for the case of $Vq\bar{q}g$ in~\cite{Gehrmann:2023zpz}. 
In summary, we can access all the production helicity amplitudes by combinations of kinematic crossings of the external momenta, parity flips and the analytic continuation of MPLs to the region denoted as (3a) in~\cite{Gehrmann:2002zr} via the substitution
\begin{align}
y = {1}/{v}\,, \quad \quad z = -{u}/{v}\,.
\end{align}

The finite remainders of the independent helicity amplitudes in the decay and production regions constitute the main result of this work and are linked in electronic format to the \code{arXiv} submission of this letter.

The provided results are compact and can easily be evaluated numerically for phenomenological studies. To demonstrate this, in table \ref{tab: numerics} we present values for the helicity coefficients at the symmetric phase space point $y=z=1-y-z=1/3$ in decay kinematics (where the coefficients $\alpha$ and $\gamma$ take the same absolute value and where all imaginary parts are generated from the expansion of the overall factors $(-q^2)^{-\epsilon}$) and at a selected point in each of the production regimes.
The values were obtained using \code{GiNaC}~\cite{Vollinga:2004sn} as implemented in \code{PolyLogTools}~\cite{Duhr:2019tlz}.

We verified that in all helicity configurations, the symbol of the leading colour $Vgq\bar{q}$ amplitude satisfies the conjectured non-adjacency conditions~\cite{Dixon:2020bbt}. It was found in~\cite{Henn:2023vbd} that in the tennis-court topology, a single master integral with 8 propagators and its crossing have a symbol with $1-z$ and $1-x$ appearing next to each other. This violation propagates to several higher topologies during integration and the cancellation of the violating symbol in the amplitude is therefore non-trivial.
\section{Conclusions}\label{sec:conclusions}
In this letter, we presented the first calculation of a planar three-loop four-point amplitude in QCD with an external mass. This calculation opens the door to a wealth of three-loop amplitudes, indispensable to precision phenomenology at hadron colliders: $V$+jet production at the LHC or three-jet production at lepton colliders, as well as the production of a Higgs boson with a jet. As the present calculation stretches current methods to their computational limit, these tasks will require a concerted effort on the completion of the non-planar integrals, the study of the special functions which appear within, as well as new ideas for IBP reduction and improvements in computer algebra for handling large intermediate expressions. In turn, the availability of three-loop amplitudes will enable precision phenomenology at N$^3$LO in QCD, which is necessary to reach the percent-level goal ahead of the High-Luminosity run at the LHC.

\section*{Acknowledgments}
This work was supported in part by the Excellence Cluster ORIGINS funded by the Deutsche Forschungsgemeinschaft (DFG, German Research Foundation) under Germany’s Excellence Strategy – EXC-2094-390783311, by the Swiss National Science Foundation (SNF) under contract 200020-204200, and by the European Research Council (ERC) under the European Union’s research and innovation programme grant agreements 949279 (ERC Starting Grant HighPHun) and 101019620 (ERC Advanced Grant TOPUP).

\newpage
\onecolumn
\appendix
\section{UV renormalization}
In this supplemental material, we collect key formulas relevant to the calculation of the three-loop helicity amplitude coefficients and to their renormalization and infrared subtraction. 
The first three $\beta$-function coefficients are
\begin{align}
    \beta_0 &= \frac{11 C_A}{6} - \frac{2 T_R N_f}{3}\,, \\
    \beta_1 &= \frac{17 C_A^2}{6} - \frac{5 C_A T_R N_f}{3} -  C_F T_R N_f\,,\\
    \beta_2 &= -\frac{205}{72} C_A C_F N_f T_R-\frac{1415}{216} C_A^2 N_f T_R+\frac{79}{108} C_A N_f^2
   T_R^2+\frac{2857 C_A^3}{432}+\frac{11}{18} C_F N_f^2 T_R^2+\frac{1}{4} C_F^2 N_f T_R\,,
\end{align}
with the QCD colour factors
\begin{equation}
    C_A = N,\quad C_F = \frac{N^2-1}{2N},\quad T_R = \frac{1}{2}\,.
\end{equation}
\section{IR subtraction}
We define the expansions
\begin{equation}
\mathbf{\Gamma}_{\text{dip}} = \sum_{l=0}^{\infty} \mathbf{\Gamma}_{l} \, \,  \Big(\frac{\alpha_s}{2 \pi} \Big)^{l+1}, \quad \quad \, \mathbf{\Gamma}' = \frac{\partial \mathbf{\Gamma}_{\text{dip}}}{\partial \log(\mu)} = \sum_{l=0}^{\infty} \mathbf{\Gamma}'_{l} \, \,  \Big(\frac{\alpha_s}{2 \pi} \Big)^{l+1}\,.
\label{DipoleExpansion}
\end{equation}
Both of the operators diagonalize in colour space for a process with three partons so we can drop the boldface notation. Substituting into \eqref{eqn: Z_operator}, we obtain
\begin{align}
    \mathcal{Z}^{(1)} &= \frac{\Gamma_{0}'}{4 \epsilon ^2} + \frac{\Gamma _0}{2 \epsilon }\\
    \mathcal{Z}^{(2)} &= \frac{\Gamma_{0}^{\prime 2}}{32 \epsilon ^4}+\frac{\Gamma'_0}{8 \epsilon
   ^3}\left(\Gamma _0-\frac{3 \beta _0}{2}\right)+\frac{1}{4 \epsilon ^2}\left(-\beta _0 \Gamma _0+\frac{\Gamma _0^2}{2}+\frac{\Gamma'
   _1}{4}\right)+\frac{\Gamma _1}{4 \epsilon }\\
    \mathcal{Z}^{(3)} &= +\frac{\Gamma_0^{\prime 3}}{384 \epsilon ^6} +\frac{\Gamma_0^{\prime 2}}{64 \epsilon ^5}\left(\Gamma _0-3\beta _0\right) +\frac{\Gamma_{0}'}{ 9\epsilon ^4}\left(-\frac{5}{4} \beta _0 \Gamma _0 +\frac{11}{9}
   \beta _0^2 +\frac{1}{4} \Gamma _0^2 +\frac{
   \Gamma_{1}'}{8}\right)\nonumber\\
   &+ \frac{1}{\epsilon ^3}\left(\frac{1}{54} \Gamma_{0}' \left(10 C_A N_f T_R-17 C_A^2+6 C_F N_f T_R\right)+\Gamma _0
   \left(\frac{\beta _0^2}{6}+\frac{\Gamma_{1}'}{32}\right)-\frac{1}{8} \beta _0 \Gamma
   _0^2-\frac{5 \beta _0 \Gamma_{1}'}{72}+\frac{\Gamma _1 \Gamma_{0}'}{16}+\frac{\Gamma
   _0^3}{48}\right)\nonumber\\
   &+\frac{1}{\epsilon ^2}\left(\Gamma _0 \left(\frac{1}{36} \left(10 C_A N_f T_R-17 C_A^2+6 C_F N_f
   T_R\right)+\frac{\Gamma _1}{8}\right)-\frac{\beta _0 \Gamma _1}{6}+\frac{\Gamma'
   _2}{36}\right)+\frac{\Gamma _2 + \Delta_3^{(3)}}{6 \epsilon }\,.
\end{align}
Defining a perturbative expansion for the anomalous dimensions with $i=K,q,\overline{q},g$\,,
\begin{align}
\gamma^{i} = \sum_{l = 0}^{\infty} \gamma_l^{i} \Big( \frac{\alpha_s}{2 \pi} \Big)^{l+1}\,,
\end{align}
we can use the dipole formula to write for the $gq\bar{q}$ system
\begin{align}
\Gamma_{l} &=  -C_F \, L_{12} \,  \gamma^{K}_l - \frac{C_A}{2} \Big(-L_{12} + L_{23} + L_{13} \Big) \gamma^{K}_l  + 2 \gamma_l^{q} + \gamma_l^{g},
\label{eqn: GammaExpansion}\\
\Gamma_{l}' &= - \gamma^{K}(\alpha_s) (2C_F+C_A)\,,
\label{eqn: gamma prime}
\end{align}
with
\begin{equation}
    L_{ij} = \log{\Big(-\frac{\mu^2}{s_{ij}+ i \eta} \Big)}\,.
\end{equation}
For convenience, we also reproduce the first three coefficients for the cusp anomalous dimension
\begin{flalign}
    \gamma^{K}_0 &=2&&\\
    \gamma^{K}_1 &=\left(\frac{67}{9}-\frac{\pi ^2}{3}\right) C_A-\frac{10 N_f}{9}&&\\
    \gamma^{K}_2 &=\left(-\frac{14 \zeta _3}{3}+\frac{10 \pi ^2}{27}-\frac{209}{54}\right) C_A N_f+\left(\frac{11 \zeta _3}{3}+\frac{11 \pi ^4}{90}-\frac{67 \pi ^2}{27}+\frac{245}{12}\right) C_A^2+\left(4 \zeta _3-\frac{55}{12}\right) C_F
   N_f-\frac{2 N_f^2}{27}\,,&&
\end{flalign}
the quark anomalous dimension
\begin{flalign}
    \gamma^{q}_0 &=-\frac{3 C_F}{2}&&\\
    \gamma^{q}_1 &=\left(\frac{13 \zeta _3}{2}-\frac{11 \pi ^2}{24}-\frac{961}{216}\right) C_A C_F+\left(\frac{65}{108}+\frac{\pi ^2}{12}\right) C_F N_f+\left(-6 \zeta _3+\frac{\pi ^2}{2}-\frac{3}{8}\right) C_F^2&&\\
    \gamma^{q}_2 &=\left(-\frac{241 \zeta _3}{54}+\frac{11 \pi ^4}{360}+\frac{1297 \pi ^2}{1944}-\frac{8659}{5832}\right) C_A C_F N_f+\left(-\frac{1}{3} \pi ^2 \zeta _3-\frac{211 \zeta _3}{6}-15 \zeta _5+\frac{247 \pi
   ^4}{1080}+\frac{205 \pi ^2}{72}-\frac{151}{32}\right) C_A C_F^2\nonumber&&\\
   &+\left(-\frac{11}{18} \pi ^2 \zeta _3+\frac{1763 \zeta _3}{36}-17 \zeta _5-\frac{83 \pi ^4}{720}-\frac{7163 \pi ^2}{3888}-\frac{139345}{23328}\right)
   C_A^2 C_F+\left(\frac{32 \zeta _3}{9}-\frac{7 \pi ^4}{108}-\frac{13 \pi ^2}{72}+\frac{2953}{432}\right) C_F^2 N_f\nonumber&&\\
   &+\left(-\frac{\zeta _3}{27}-\frac{5 \pi ^2}{108}+\frac{2417}{5832}\right) C_F N_f^2+\left(\frac{2 \pi
   ^2 \zeta _3}{3}-\frac{17 \zeta _3}{2}+30 \zeta _5-\frac{\pi ^4}{5}-\frac{3 \pi ^2}{8}-\frac{29}{16}\right) C_F^3\,,&&
\end{flalign}
and the gluon anomalous dimension
\begin{flalign}
    \gamma^{g}_0 &=-\beta _0&&\\
    \gamma^{g}_1 &=\left(\frac{32}{27}-\frac{\pi ^2}{36}\right) C_A N_f+\left(\frac{\zeta _3}{2}+\frac{11 \pi ^2}{72}-\frac{173}{27}\right) C_A^2+\frac{C_F N_f}{2}&&\\
    \gamma^{g}_2 &=\left(-\frac{19 \zeta _3}{9}-\frac{\pi ^4}{90}-\frac{\pi ^2}{24}+\frac{1217}{216}\right) C_A C_F N_f+\left(-\frac{7 \zeta _3}{27}+\frac{5 \pi ^2}{324}-\frac{269}{11664}\right) C_A N_f^2-\frac{11}{72} C_F N_f^2-\frac{1}{8} C_F^2 N_f\nonumber&&\\
    &+\left(\frac{89 \zeta
   _3}{54}+\frac{41 \pi ^4}{1080}-\frac{599 \pi ^2}{1944}+\frac{30715}{11664}\right) C_A^2 N_f+\left(-\frac{5}{18} \pi ^2 \zeta _3+\frac{61 \zeta _3}{12}-2 \zeta _5-\frac{319 \pi ^4}{2160}+\frac{6109 \pi
   ^2}{3888}-\frac{48593}{2916}\right) C_A^3\,.&&
\end{flalign}

\section{Projectors onto helicity amplitude coefficients}
With the common denominator $\mathcal{D} = 2(d-3)y^3z^2(1-y-z)^2$ and the tensor basis defined as in ref.~\cite{Gehrmann:2023zpz},
\begin{flalign}
\mathcal{P}_{\alpha_1} &=\frac{1}{\mathcal{D}}\Big[-d (1-y)^2 z \:T^{\dagger }_1+z (-2 y (-1+y+2 z)+d (-1+y+z+y z)) \:T^{\dagger }_2-(1-y) y z (1-y-z) \:T^{\dagger }_3\nonumber &&\\
&\phantom{=\frac{1}{\mathcal{D}}\Big[\:}+(1-y)^2 y z \:T^{\dagger }_4-y z (-1+y+z+y z)
   \:T^{\dagger }_5+(1-y) y z^2 \:T^{\dagger }_6\Big]\,,&&\\
\mathcal{P}_{\alpha_2} &=\frac{1}{\mathcal{D}}\Big[-z \left(-(d-4) y^2+(d-4) y (1-z)+d z\right) \:T^{\dagger }_1+z \left(2 y^2-d (1-z) z+y (-4+d+(2+d) z)\right) \:T^{\dagger
   }_2\nonumber &&\\
&\phantom{=\frac{1}{\mathcal{D}}\Big[\:}-y z (1-y-z) (y+z) \:T^{\dagger }_3-y z (z+y (1-y-z)) \:T^{\dagger }_4-y z \left(y-(1-y) z+z^2\right) \:T^{\dagger }_5+y z^2 (y+z) \:T^{\dagger }_6\Big]\,,&&\\
\mathcal{P}_{\alpha_3} &=\frac{1}{\mathcal{D}}\Big[+4 (1-y) y z
   (1-y-z) \:T^{\dagger }_1+4 y z (1-y-z)^2 \:T^{\dagger }_2+2 y^2 z (1-y-z)^2 \:T^{\dagger }_3-2 y^2 z^2 (1-y-z) \:T^{\dagger }_6\Big]\,, &&\\
\mathcal{P}_{\gamma_1} &=\frac{1}{\mathcal{D}}\Big[+(d-4) y (-1+y+z+y z)
   \:T^{\dagger }_1+y (1-z) (-4+d+2 y+4 z-d z) \:T^{\dagger }_2-y^2 (1-z) (1-y-z) \:T^{\dagger }_3\nonumber &&\\
&\phantom{=\frac{1}{\mathcal{D}}\Big[\:}-y^2 (-1+y+z+y z) \:T^{\dagger }_4+y^2 (1-z)^2 \:T^{\dagger
   }_5-y^2 (1-z) z \:T^{\dagger }_6\Big]\,, &&\\
\mathcal{P}_{\gamma_2} &=\frac{1}{\mathcal{D}}\Big[+(d-4) y (z-y (1-y-z)) \:T^{\dagger }_1+y \left(2 y^2+(d-4) (1-z) z+y (-4+d-(d-2) z)\right) \:T^{\dagger }_2\nonumber &&\\
&\phantom{=\frac{1}{\mathcal{D}}\Big[\:}-y^2
   (1-y-z) (y+z) \:T^{\dagger }_3-y^2 (z-y (1-y-z)) \:T^{\dagger }_4+y^2 \left(y-(1-y) z+z^2\right) \:T^{\dagger }_5-y^2 z (y+z) \:T^{\dagger }_6\Big]\,, &&\\
\mathcal{P}_{\gamma_3} &=\frac{1}{\mathcal{D}}\Big[+4 y^2 z (1-y-z)
   \:T^{\dagger }_2-2 y^2 z (1-y-z)^2 \:T^{\dagger }_3-2 y^2 z^2 (1-y-z) \:T^{\dagger }_6\Big]\,.
\end{flalign}

\twocolumn

\bibliographystyle{elsarticle-harv} 
\bibliography{main}

\end{document}